\documentclass[journal,transmag]{IEEEtran}
\usepackage[caption=false,font=normalsize,labelfont=sf,textfont=sf]{subfig}
\usepackage{graphicx,color,epsf,bm}
\usepackage{amsmath,amsfonts,amssymb,amscd,bm}
\usepackage{flushend}
\usepackage[none]{hyphenat}
\usepackage{tikz}
\newcommand{\Y}{\mathbf{Y}}
\newcommand{\y}{\mathbf{y}}

\newcommand{\E}{\mathrm{E}}
\newcommand{\Cov}{\mathrm{Cov}}
\newcommand{\Var}{\mathrm{Var}}
\newcommand{\C}{\mathbf{C}}
\renewcommand{\Re}[1]{\mathrm{Re}(#1)}
\renewcommand{\Im}[1]{\mathrm{Im}(#1)}

\usepackage{eso-pic}
\AddToShipoutPicture*{\footnotesize\sffamily\raisebox{0.6cm}{\hspace{1.65cm}\fbox{\parbox{\textwidth}{\copyright~2017 IEEE. Personal use of this material is permitted. Permission from IEEE must be obtained for all other uses, in any current or future media, including reprinting/republishing this material for advertising or promotional purposes, creating new collective works, for resale or redistribution to servers or lists, or reuse of any copyrighted component of this work in other works.}}}}

\begin{document}
\title{Low-Dimensional Stochastic Modeling of the Electrical Properties of Biological Tissues}
\author{\IEEEauthorblockN{Ulrich R\"omer\IEEEauthorrefmark{1}$^,$\IEEEauthorrefmark{2},
Christian Schmidt\IEEEauthorrefmark{3},
Ursula van Rienen\IEEEauthorrefmark{3}, and
Sebastian Sch\"ops\IEEEauthorrefmark{1}$^,$\IEEEauthorrefmark{2}}\newline
\IEEEauthorblockA{\IEEEauthorrefmark{1}Technische Universit\"at Darmstadt, Institut f\"ur Theorie Elektromagnetischer Felder, Darmstadt, Germany}\newline
\IEEEauthorblockA{\IEEEauthorrefmark{2}Technische Universit\"at Darmstadt, Graduate School of Computational Engineering, Darmstadt, Germany}\newline
\IEEEauthorblockA{\IEEEauthorrefmark{3}Universit\"at Rostock, Institut f\"ur Allgemeine Elektrotechnik, Rostock, Germany}
\thanks{Manuscript received xxx; revised xxx. 
Corresponding author: U. R\"omer (email: roemer@temf.tu-darmstadt.de).}}

\maketitle

\begin{abstract}
Uncertainty quantification plays an important role in biomedical engineering as measurement data is often unavailable and literature data shows a wide variability. Using state-of-the-art methods one encounters difficulties when the number of random inputs is large. This is the case, e.g., when using composite Cole-Cole equations to model random electrical properties. It is shown how the number of parameters can be significantly reduced by the Karhunen-Lo\`{e}ve expansion. The low-dimensional random model is used to quantify uncertainties in the axon activation during deep brain stimulation. Numerical results for a Medtronic 3387 electrode design are given.
\end{abstract}

\begin{IEEEkeywords}
Uncertainty, random processes, principal component analysis, biomedical engineering.
\end{IEEEkeywords}


\section{Introduction}
\IEEEPARstart{T}{he} electrical properties of biological tissue are based on experimental data and are subject to large variability in literature \cite{gabriel2009, schmidtieee2013}, which arises from difficulties associated with the measuring process. Their properties vary over frequency and exhibit a non-symmetrical distribution of relaxation times, which can be described by composite Cole-Cole equations. Randomness in the material can be accounted for by modeling the parameters in the Cole-Cole equations as random variables. This gives rise to random material laws which are physically motivated but contain a large number of random parameters. Hence, they are not well suited for the majority of uncertainty quantification methods that scale unfavorably with the dimension of the parameter space. 

In this study we exploit correlation in the random Cole-Cole equation to substantially reduce the number of parameters. In particular, we use an eigendecomposition of the covariance matrix to derive a low-rank approximation. The truncated Karhunen-Lo\`{e}ve (KL) expansion \cite{loeve1978,ghanem1991} of the random material is then spanned in direction of the dominant eigenfunctions. This procedure is closely related to principal component analysis and proper orthogonal decomposition.

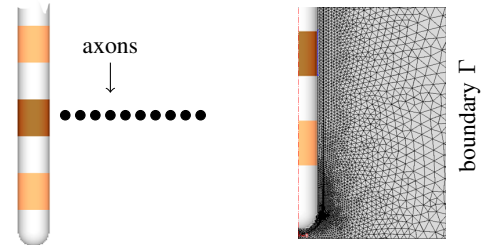
\begin{figure}
\begin{minipage}[t!]{0.2\textwidth}
\centering
\begin{tikzpicture}	
	\node at (0,0) {\includegraphics[width=0.13\columnwidth]{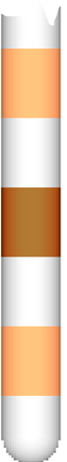}};	
	\draw[black,fill=black] (0.4,0.1) circle (.4ex);
	\draw[black,fill=black] (0.6,0.1) circle (.4ex);
	\draw[black,fill=black] (0.8,0.1) circle (.4ex);
	\draw[black,fill=black] (1,0.1) circle (.4ex);
	\draw[black,fill=black] (1.2,0.1) circle (.4ex);
	\draw[black,fill=black] (1.4,0.1) circle (.4ex);
	\draw[black,fill=black] (1.6,0.1) circle (.4ex);
	\draw[black,fill=black] (1.8,0.1) circle (.4ex);
	\draw[black,fill=black] (2,0.1) circle (.4ex);
	\draw[black,fill=black] (2.2,0.1) circle (.4ex);
	\node at (1,1) {\small axons};
	\draw[->] (1,0.8) -- (1,0.4);
\end{tikzpicture}
\end{minipage}
\begin{minipage}[t!]{0.2\textwidth}
\centering
\begin{tikzpicture}	
	\node at (0,0) {\includegraphics[width=0.54\columnwidth]{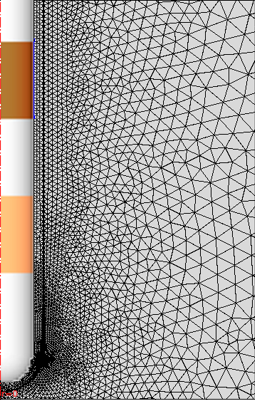}};	
	\node[rotate=90] at (1.3,0.1) {\small boundary $\Gamma$};
\end{tikzpicture}
\end{minipage}
\caption{Axons aligned perpendicular to electrode (left) and computational domain with mesh using rotational symmetry (right)}
\label{fig:axons}
\end{figure}

The final computational goal is to quantify uncertainties in the axon activation during Deep Brain Stimulation (DBS) \cite{schmidtieee2013}. To this end, the stimulation electrode and the surrounding brain tissue are modeled as a volume conductor, see Figure \ref{fig:axons} (left). A numerical approximation of the electric potential is obtained by the finite element method. The quantity of interest is the minimal electrode current to be applied in order to activate a particular axon in the electrode's vicinity. This optimization is formulated as a root-finding problem for a function obtained from post-processing the solution of the volume conductor problem. Brent's method is applied for its numerical solution. 

\vspace*{0.5em}

In the presence of randomness in the electric coefficients, uncertainty quantification techniques are required. We use a stochastic quadrature on sparse grids \cite{xiu2005,babuska2010} to efficiently compute the mean value and standard deviation of the axon activation. The method is non-intrusive as it only requires repetitive runs of the volume conductor model and the activation potential post-processing routine. 

The paper is organized as follows: Sections \ref{sec:cole} and \ref{sec:KL} contain the random Cole-Cole equation together with the KL expansion. Section \ref{sec:problem} briefly summarizes the main equations needed for modeling DBS. Section \ref{sec:uq} introduces a stochastic setting together with the stochastic quadrature. Finally, numerical results for a Medtronic 3387 electrode design are given in Section \ref{sec:num}.

\section{Random Cole-Cole Equation}
\label{sec:cole}
Electrical properties of biological tissues can be modeled by the Cole-Cole equation 
\begin{equation}
f(\omega) =  \epsilon_\infty + \frac{\varkappa_i}{j \omega  \epsilon_0} + \sum_{i=1}^4 \frac{\Delta \epsilon_n}{1+(j \omega \tau_n)^{1-\alpha_n}}, 
\label{eq:cole_cole}
\end{equation}
where $\omega$ denotes frequency, $j$ the imaginary unit, $\varkappa_i$ the static ionic conductivity and $\tau_n$ represents relaxation time constants. Also, $\epsilon_\infty$ and $\Delta \epsilon_n$ denote the high frequency and difference of the low to high frequency relative permittivitiy, respectively. From \eqref{eq:cole_cole} the permittivity and electric conductivity are inferred as $\epsilon(\omega) = \Re{f(\omega)}$ and $\varkappa(\omega) = -\Im{\epsilon_0 \omega f(\omega)}$, with $\mathrm{Re}$ and $\mathrm{Im}$ referring to the real part and imaginary part, respectively.

In \eqref{eq:cole_cole}, $\epsilon_\infty,\varkappa_i,\Delta \epsilon_n,\tau_n$ and $\alpha_n$ are parameters that need to be inferred from measurements. As uncertainties are inevitably connected to this process we consider these parameters to be random variables $Y_i: \Theta \rightarrow \mathbb{R}$, $i=1,\ldots,14$, where $\Theta$ refers to a set of random outcomes. Then, with $\theta \in \Theta$ denoting a random event, the random Cole-Cole equation reads
\begin{equation}
f(\theta,\omega) =  Y_1(\theta) + \frac{Y_2(\theta)}{j \omega  \epsilon_0} + \sum_{i=1}^4 \frac{Y_{3 i}(\theta)}{1+(j \omega Y_{3 i + 1}(\theta))^{1-Y_{3 i + 2}(\theta)}}.
\label{eq:random_cole_cole}
\end{equation}

In view of \eqref{eq:random_cole_cole}, both the electric permittivity and the conductivity are random. Since the following derivation is identical for both $\epsilon$ and $\varkappa$, we use the function $g$ referring to either of them. 
Important measures of the random field $g$ are the expected value and the covariance, given as
\begin{align}
\E_g(\omega) &= \int_{\Theta} g(\theta,\omega) \ \mathrm{d} P(\theta), \\
\Cov_g(\omega,\omega') &= \int_{\Theta} (g(\theta,\omega)-\E[g](\omega)) \notag \\
  & \hspace*{3em} \cdot (g(\theta,\omega')-\E[g](\omega')) \ \mathrm{d} P(\theta),
\end{align}
where $P$ refers to a probability measure.

\section{Discrete Karhunen-Lo\`{e}ve Expansion}
\label{sec:KL}
When \eqref{eq:random_cole_cole} is used within simulations, both the large number of random variables and their possible correlation pose difficulties. The former results in a high computational complexity, whereas a possible correlation of the inputs cannot be handled by many state-of-the-art uncertainty quantification methods. In the following we apply the discrete Karhunen-Lo\`{e}ve expansion (KLE) to reduce the number of random variables in \eqref{eq:random_cole_cole}. Although, the KLE is readily applicable to random fields such as \eqref{eq:random_cole_cole}, we consider its discrete variant, also referred to as principal component analysis. The exposition thereby follows \cite{elia2013coarse}.

Given a set of frequency points $\{\omega_n\}_{n=1}^N$, chosen equidistantly over a fixed interval on a logarithmic scale, we consider the covariance matrix $\C$ with entries
\begin{equation}
C_{n_1,n_2} = \Cov_g(\omega_{n_1},\omega_{n_2}), \ n_1,n_2=1,\ldots,N
\label{eq:cov}
\end{equation}
and denote its eigenvectors and eigenvalues with $\mathbf{b}_n$ and $\lambda_n$, respectively. Then, $\C$ can be decomposed as 
\begin{equation}
\mathbf{C} = \mathbf{V} \mathbf{E} \mathbf{V}^\top,
\end{equation}
with $\mathbf{V}$ storing the eigenvectors $\mathbf{b}_n$ column-wise and $\mathbf{E}$ containing the eigenvalues $\lambda_n$ in decreasing order on its diagonal. As $\C$ is symmetric positive definite, the $\lambda_n$ are real and positive. Moreover, given a strongly correlated random field $g$, the eigenvalues decrease rapidly \cite{elia2013coarse}. Hence, we only consider the $M$ largest eigenvalues by introducing 
\begin{equation}
\mathbf{C} \approx \mathbf{C}_M = \mathbf{V}_M \mathbf{E}_M \mathbf{V}_M^\top.
\label{eq:rank_M} 
\end{equation}
Numerical examples discussing the error committed by this low-rank approximation are given in Section \ref{sec:num}. As only the largest eigenvalues and eigenfunctions are required a Krylov subspace method, such as the Lanczos algorithm can be used. Moreover, the underlying random field has only one dimension (frequency). This results in a moderate size of the covariance matrix and acceleration techniques for the matrix-vector product can be omitted. Numerical techniques and properties of the Karhunen-Lo\`{e}ve expansion for one-dimensional random fields were also investigated in \cite{roemer2016} in the context of nonlinear magnetic material properties.  

Based on \eqref{eq:rank_M} a new discrete random field is defined as
\begin{equation}
\mathbf{g}_{M}(\theta) =  \E_{\mathbf{g}} + \mathbf{V}_M \mathbf{E}_M^{1/2} \Y_M(\theta), 
\label{eq:KL}
\end{equation}
where $(\E_{\mathbf{g}})_n = \E_g(\omega_n)$. A frequency dependent random field $g_M$ is recovered from $\mathbf{g}_{M}$ by spline interpolation. The new random variables are uncorrelated and can be inferred from
\begin{equation}
Y_{M;m}(\theta) =  \left((\mathbf{g}(\theta) - \E_{\mathbf{g}} )^\top \mathbf{b}_m \right )/\sqrt{\lambda_m}, \ m=1,\dots,M,
\end{equation}
based on observations $(\mathbf{g}(\theta))_n= g(\theta,\omega_n)$. It should be noted that the variables $\Y_M$ are also independent in the case of a Gaussian random field. In general independence needs to be assured by introducing a transformation to another set of random variables. Here, we simply assume independence. \section{Problem Description}
\label{sec:problem}
The computational model to estimate the activation during DBS is based on a 2D rotational symmetric finite element volume conductor model of the stimulation electrode and surrounding brain tissue coupled to axons in the target area. The electric potential in the tissue is computed by solving the Laplace equation for complex material properties
\begin{equation}
\nabla \cdot \left[\left(\varkappa(\omega, \boldsymbol r)+\mathsf{j}\omega\epsilon_0\epsilon_r(\omega, \boldsymbol r)\right) \nabla \phi_{\mathrm{e}}(\omega, \boldsymbol r)\right] = 0
\label{eq:potential}
\end{equation}
with the electric conductivity $\varkappa(\omega, \boldsymbol r)$ and relative permittivity $\epsilon_r(\omega, \boldsymbol r)$ of the encapsulation layer and brain tissue. Following the approach in \cite{schmidtieee2013}, a current-controlled stimulation pulse $I(\omega)$ is introduced to one electrode contact, while the other boundaries of the electrode are modeled as insulation. The boundary of the surrounding tissue is set to ground, i.e., $\phi_\mathrm{e} |_{\Gamma} = 0$, see Figure \ref{fig:axons} (right).

The time-dependent electric potential resulting from the applied stimulus is computed using the Fourier Finite Element Method (FFEM) \cite{butson2006}, for which the Laplace equation (\ref{eq:potential}) is solved in the frequency-domain for $N$ logarithmically distributed frequency nodes and interpolated for the Fourier components of the stimulation signal in the considered frequency range. 
In order to investigate the activation of neuronal tissue during DBS, a number of axon cable models are positioned perpendicular to the electrode contact where the stimulus is applied. Each axon cable model consists of a number of compartments, for which the inner potential in each compartment is defined by the following equation \cite{mcintyre2002}:
\begin{align}
\begin{split}
g_\mathrm{A}(\boldsymbol r)\Delta^2\phi_\mathrm{e}(\boldsymbol r, t)&=c(\boldsymbol r) \frac{\mathrm{d}\phi_\mathrm{m}(\boldsymbol r, t)}{dt}+\\
&+i_\mathrm{ion}(\phi_\mathrm{m}(\boldsymbol r, t),\boldsymbol r)-\\
&-g_\mathrm{A}(\boldsymbol r)\Delta^2\phi_\mathrm{m}(\boldsymbol r, t)
\end{split}
\label{eq:neuron}
\end{align}
with the membrane capacitance $c$, the ionic current $i_\mathrm{ion}$, the axial conductance $g_\mathrm{A}$, the membrane potential $\phi_\mathrm{m}(\boldsymbol r, t)$, and second spatial difference $\Delta^2$ in direction of the axon. The membrane potential is defined by
\begin{equation}
\phi_\mathrm{m}(\boldsymbol r, t)=\phi_\mathrm{i}(\boldsymbol r,t)-\phi_\mathrm{e}(\boldsymbol r,t)+\phi_\mathrm{r}(\boldsymbol r,t)
\label{eq:membranepotential}
\end{equation}
with the innercellular potential $\phi_\mathrm{i}(\boldsymbol r,t)$, the resting potential $\phi_\mathrm{r}(\boldsymbol r,t)$, and the extracellular potential $\phi_\mathrm{e}(\boldsymbol r,t)$. The time-dependent electric potential at each compartment center provided by the volume conductor model is applied as extracellular potential $\phi_\mathrm{e}(\boldsymbol r, t)$ to the compartment equation~(\ref{eq:neuron}). 

The computational goal is to determine the minimum stimulation amplitude, required to excite an action potential in a specific axon. This can be expressed as a root-finding problem as follows: for a given current stimulus, problem \eqref{eq:potential} is solved repeatedly as outlined above to obtain a time-dependent potential $\phi_\mathrm{e}$. Then, for the axon under investigation, for each compartment, \eqref{eq:neuron} is solved. The axon is activated, if the inner potential at the outer compartment $\phi_\mathrm{i}^{\mathrm{out}}$ is larger than zero. Hence, to determine the required stimulation amplitude $I$, a root of $\phi_\mathrm{i}^{\mathrm{out}}(I)$ needs to be found. This root is found numerically with Brent's method here.  
 \section{Uncertainty Quantification}
\label{sec:uq}
Randomness in the electric conductivity and relative permittivity gives rise to a stochastic volume conductor model. This stochastic model in turn can be used to compute statistics of the axon activation current $I$. The methods presented in this section assume independence of the inputs. Unfortunately, the KLE as presented in Section \ref{sec:KL} applied to both the permittivity and the conductivity does not ensure independence of the parameters, as both are modeled by the same random process~\eqref{eq:random_cole_cole}. An extension to obtain uncorrelated and independent electrical parameters at the same time is possible but beyond the scope of this paper. We have observed that the conductivity is the parameter with a larger sensitivity and hence, in the following, only $\varkappa$ is subject to uncertainty.
We obtain the parametric equation
\begin{align}
\!\!\nabla \cdot \left[\left(\varkappa(\y_{M},\omega,\boldsymbol r)+\mathsf{j}\omega\epsilon_0\epsilon_r(\omega,\boldsymbol r)\right) \nabla \phi_{\mathrm{e}}(\y_M,\omega, \boldsymbol r)\right] = 0,
\label{eq:vol_cond_parametric}
\end{align}
where lowercase symbols are used for the realization of a random variable, i.e., $\y_M =\Y_M(\theta)$. The parameter dependency is inherited by the activation potential and the minimum current required for activation. In particular, a current is associated to each $\y_M$ through
\begin{equation}
\phi_\mathrm{i}^{\mathrm{out}}(\y_M,I) = 0.
\label{eq:root_para}
\end{equation}

We denote with $\rho$ and $\Gamma$ the joint probability density function and the image of $\Y_M$, respectively. Then the expected value and variance can be rewritten as
\begin{subequations}
\begin{align}
\E[I] &= \int_{\Gamma} I(\y_M) \ \rho(\y_M) \mathrm{d} \y_M, \\
\Var[I] &= \int_{\Gamma} (I(\y_M) - \E[I])^2 \ \rho(\y_M) \mathrm{d} \y_M.
\end{align}
\label{eq:moments}
\end{subequations}
The aim is to find an efficient numerical approximation of~\eqref{eq:moments}.

A state-of-the art technique for uncertainty quantification is the stochastic collocation method \cite{xiu2005,babuska2010} based on tensor or sparse grids. The procedure is summarized as follows: given a set of collocation points $(\y_M^{(k)})_{k=1}^K$, \eqref{eq:root_para} is solved for each $\y_M^{(k)}$ to obtain $I(\y_M^{(k)})$. This involves solutions of the volume conductor model and post-processing to obtain the respective action potentials. We emphasize that no modification of the code is required as simulations are simply repeated with different conductivities $\varkappa(\y_M^{(k)})$. In this sense, the method is non-intrusive. The collocation points are given by tensor grid or sparse grid constructions. 
 
Given $(I(\y_M^{(k)}))_{k=1}^K$, a polynomial approximation of the output quantity can be computed by enforcing the collocation conditions. Here, we are mainly interested in the approximation of the expected value and variance which can be directly obtained using a dedicated numerical quadrature. The knots and weights of univariate quadrature rules are given for instance by the Gauss or Clenshaw-Curtis abscissas. Then, tensor product formulas of different degree in different directions are combined to obtain efficient quadrature rules, see, e.g., \cite{novak1999}, \cite{babuska2010}. \section{Numerical Example}
\label{sec:num}
In a first step, the KLE was applied to the random conductivity given by \eqref{eq:random_cole_cole}. The random vector $\Y$ was modeled with a mean value 
\begin{multline}
\E[\Y]= {\scriptstyle(4,0.02,45,7.96 \times10^{-12},0.1,400,15.92\times 10^{-9},0.15, }\\  {\scriptstyle2 \times 10^5, 106.10 \times 10^{-6},0.22,4.5 \times 10^{7},5.31 \times 10^{-3},0)^\top}
\end{multline}
according to values given in literature. Due to the lack of further data, the vector was assumed to be uniformly distributed on a interval of 10$\%$ deviation around the mean value. Equidistant points with a stepsize of $0.004$ on the logarithmic scale of the interval $2 \pi [130, 5 \cdot 10^5]$ Hz were considered. The eigenvalues and eigenfunctions were computed with the eigs function of MATLAB. 

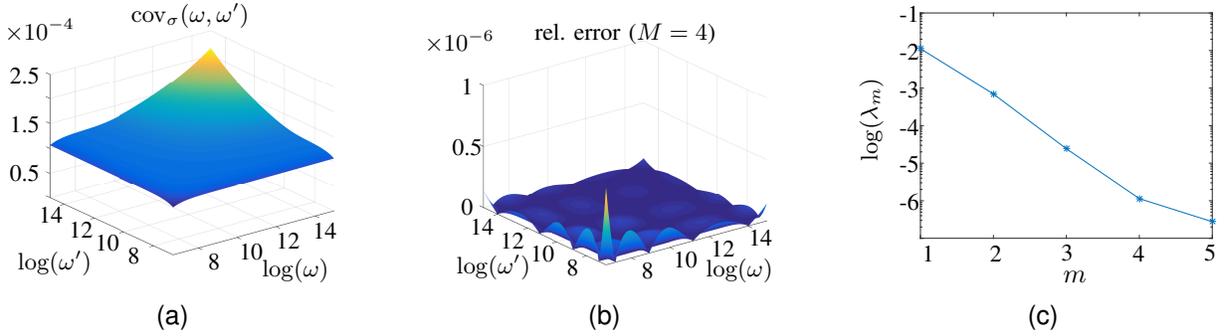
\begin{figure*}[!t]
\centering
\subfloat[]{
\begin{tikzpicture}	
	\node at (-2.2,0) {\includegraphics[width=0.45\columnwidth]{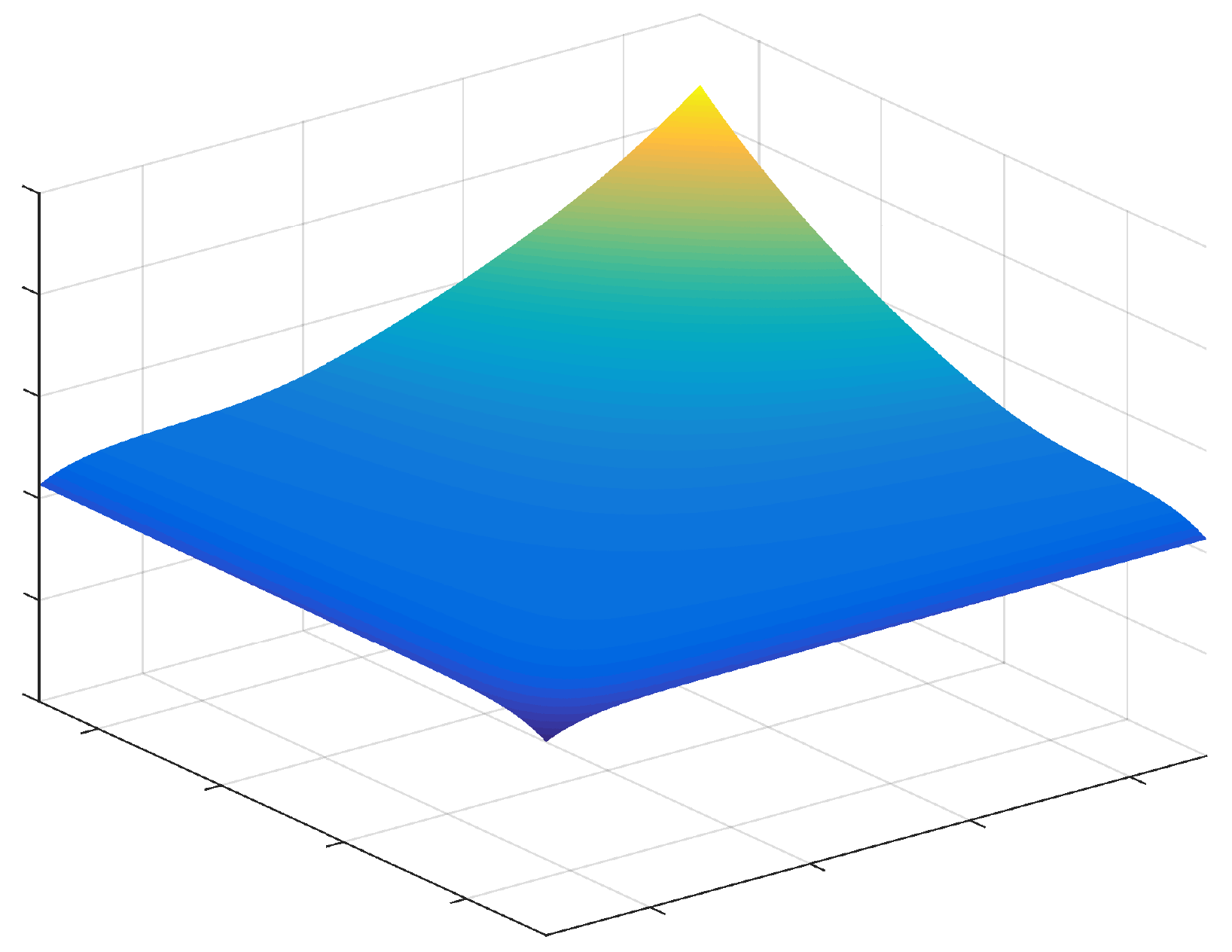}};
	\node at (-0.8,-1.7) {\small $\log(\omega)$};
	\node at (-4,-1.6) {\small $\log(\omega')$};
	\node at (-2.2,1.7) {\small $\mathrm{cov}_\sigma(\omega,\omega')$};
	\node at (-0.41,-1.16) {\small 14};
	\node at (-0.9,-1.3) {\small 12};
	\node at (-1.39,-1.44) {\small 10};
	\node at (-1.88,-1.58) {\small 8};
	\node at (-2.95,-1.5) {\small 8};
	\node at (-3.3,-1.32) {\small 10};		
	\node at (-3.65,-1.15) {\small 12};	
	\node at (-4.05,-0.95) {\small 14};	
	\node at (-4.4,-0.4) {\small 0.5};	
	\node at (-4.4,0.25) {\small 1.5};	
	\node at (-4.4,0.9) {\small 2.5};	
	\node at (-4.2,1.5) {\small $\times 10^{-4}$};		
\end{tikzpicture}
\label{fig:cov_a}}
\hfil
\subfloat[]{\begin{tikzpicture}	
	\node at (2.2,0) {\includegraphics[width=0.45\columnwidth]{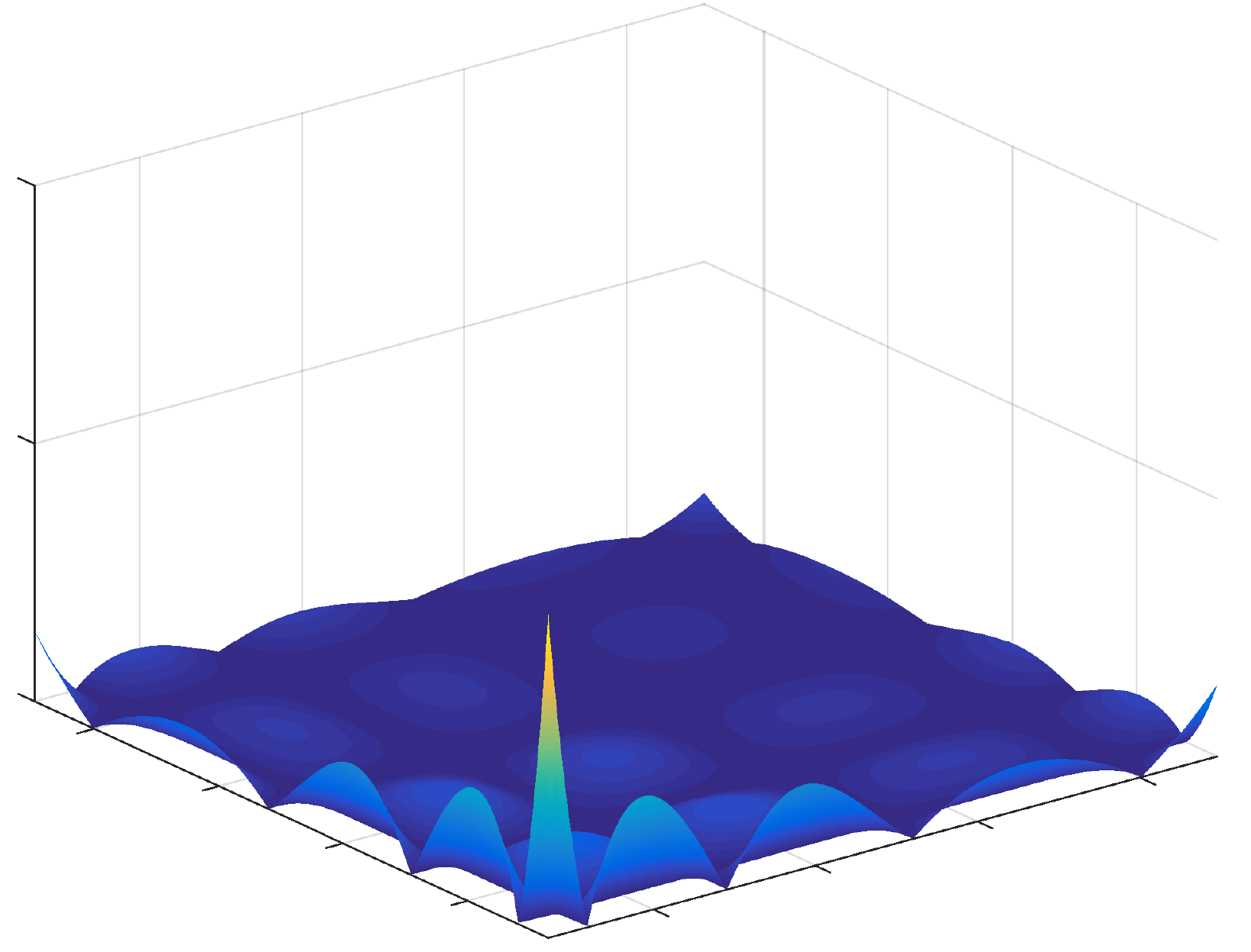}};
	\node at (0.45,-1.5) {\small $\log(\omega')$};
	\node at (3.72,-1.55) {\small $\log(\omega)$};
	\node at (2.2,1.6) {\small rel. error ($M=4$)};	
	\node at (0.3,-0.96) {\small 14};
	\node at (0.7,-1.14) {\small 12};
	\node at (1.1,-1.32) {\small 10};
	\node at (1.5,-1.5) {\small 8};
	\node at (2.5,-1.55) {\small 8};	
	\node at (3,-1.4) {\small 10};	
	\node at (3.5,-1.25) {\small 12};	
	\node at (4,-1.1) {\small 14};	
	\node at (0.1,-0.7) {\small 0};	
	\node at (0.1,0.1) {\small 0.5};	
	\node at (0.1,0.9) {\small 1};	
	\node at (0,1.5) {\small $\times 10^{-6}$};	
\end{tikzpicture}
\label{fig:cov_b}}
\hfil
\subfloat[]{\begin{tikzpicture}	
	\node at (2.2,0) {\includegraphics[width=0.45\columnwidth]{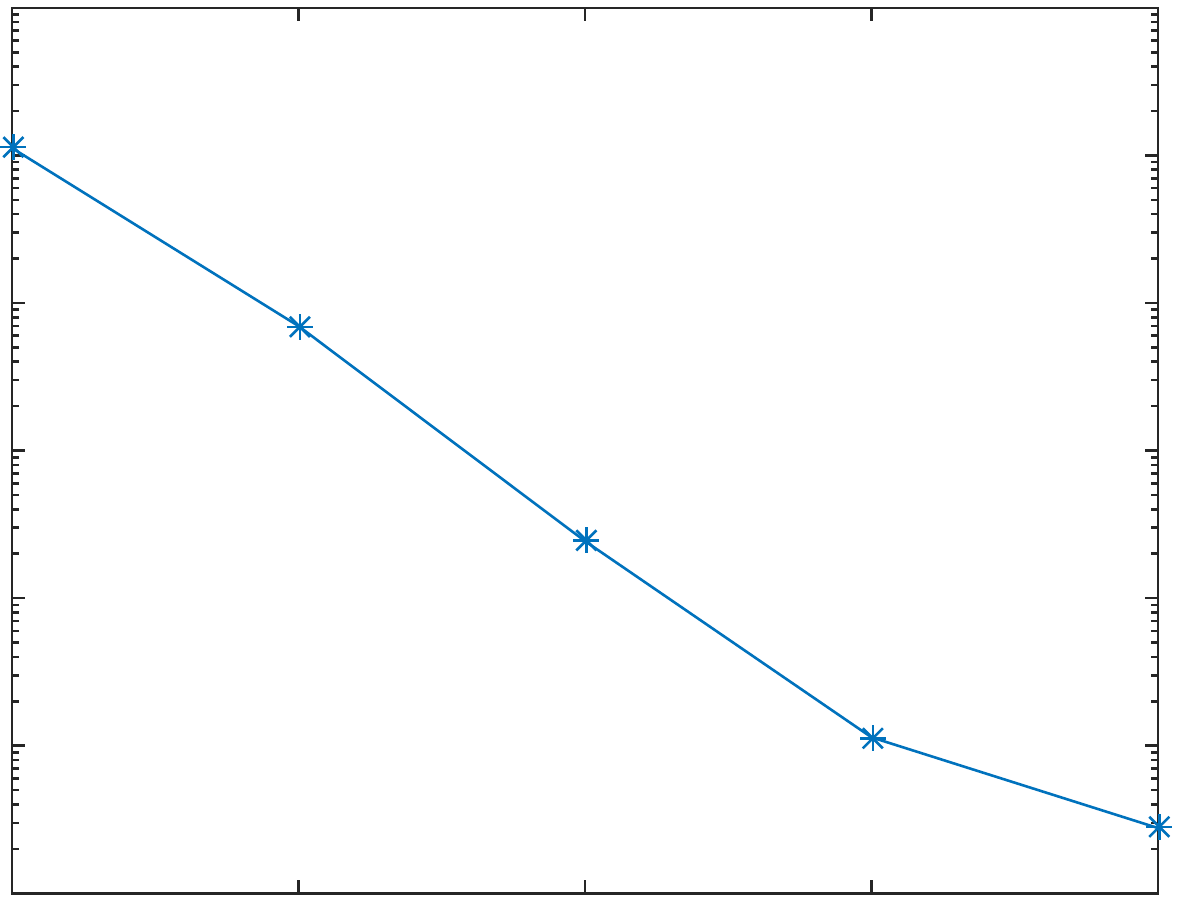}};
	\node[rotate=90] at (-0.4,0.1) {\small $\log(\lambda_m)$};	
	\node at (2.3,-2) {$m$};	
	\node at (0.35,-1.68) {\small 1};
	\node at (1.25,-1.68) {\small 2};
	\node at (2.2,-1.68) {\small 3};
	\node at (3.15,-1.68) {\small 4};
	\node at (4.1,-1.68) {\small 5};
	\node at (0.1,-1) {\small -6};	
	\node at (0.1,-0.5) {\small -5};	
	\node at (0.1,0.0) {\small -4};	
	\node at (0.1,0.5) {\small -3};	
	\node at (0.1,1) {\small -2};	
	\node at (0.1,1.5) {\small -1};	
\end{tikzpicture}
\label{fig:cov_c}}
\caption{(a) Sample covariance of random conductivity modeled by \eqref{eq:random_cole_cole} (b) Relative error of Karhunen-Lo\`{e}ve expansion and $M=4$ (c) Decay of eigenvalues}
\label{fig:cov}
\end{figure*}

Figure \ref{fig:cov}(a) depicts the covariance sampled $10^{3}$ times. The relative error for the KL expansion and $M=4$ is shown in Figure \ref{fig:cov}(b). As it is in the order of $10^{-6}$ the low-rank approximation is justified. This is further illustrated in Figure \ref{fig:cov}(c) showing the fast (exponential) decay of the eigenvalues. 

The low-dimensional stochastic model for the conductivity was then used for the simulation of the required axon activation current. Simulation details are summarized as follows: the electrode model, which represents the Medtronic 3387 electrode design commonly used in human DBS \cite{schmidtieee2013}, was encapsulated by a $0.2\,\mathrm{mm}$ thick tissue layer, which is formed due to body reactions at the interface between the electrode and the brain tissue \cite{grant2010}. Cathodal current-controlled square-wave stimulation pulses with a frequency of $130\,\mathrm{Hz}$ and a pulse duration of $60\,\mathrm{\mu s}$, as used in clinical practice \cite{schmidtieee2013}, were applied. The model equation \eqref{eq:potential} was discretized with $27{,}000$ elements and solved using the software COMSOL Multiphysics\textregistered $\ $ at $N=3846$ frequency points in the considered frequency range. By subsequent refinement of the frequency interval an accuracy of 1$\%$ was ensured. Ten axon cable models, each of them with 221 compartments, were positioned perpendicular to the second electrode contact in a distance between $1\,\mathrm{mm}$ to $10\,\mathrm{mm}$ to the electrode center. The axon activation was obtained by solving \eqref{eq:neuron} with the backward Euler method. A time step of $10\,\mathrm{\mu s}$ was employed. The minimal current required for axon activation was computed with Brent's method with an absolute tolerance of $1\cdot10^{-5}$.

Table \ref{tab:uq} gives the mean value and standard deviation of the axon activation obtained with a stochastic quadrature. Clenshaw-Curtis abscissas were used for the univariate quadrature rules. Quadrature points and weights for the multivariate case were chosen to exactly integrate total degree polynomials of level three, which corresponds to 137 quadrature points and weights in total. Details can be found in \cite{babuska2010}. The values in Table \ref{tab:uq} are rounded to significant figures, estimated with a higher order quadrature. It is observed that the standard deviation is in the order of 10$\%$ of the mean value for each axon. This reflects a moderate sensitivity of the goal fucntion with respect to the variable conductivity input parameters.
\begin{table}[t!]
\label{tab:uq}
\centering
\caption{Expected value and standard deviation of axon activation using stochastic quadrature}
\begin{tabular}{|c||c|c|}
\hline
Axon & mean value [mA]& standard deviation [mA]  \\
\hline
\hline
1 &  0.14 & 0.01 \\
\hline
2 &  0.56 & 0.06\\
\hline
3 &  1.44 & 0.15\\
\hline
4 &  2.97 & 0.31\\ 
\hline
5 &  5.30 & 0.56\\
\hline
6 &  8.65 & 0.91\\
\hline
7 &  13.20 & 1.39\\
\hline
8 &  19.15 & 2.02\\
\hline
9 &  26.67 & 2.81\\
\hline
10 &  36.06 & 3.80\\
\hline
\end{tabular}
\end{table}

\section{Conclusion}
Uncertainties in the axon activation in deep brain stimulation have been quantified using a stochastic quadrature and a volume conductor model for the electric potential distribution in the brain tissue. The activation was found to be moderately sensitive to variations in the conductivity parameters. Hence, the problem is well conditioned with respect to deviations in the electrical input parameters. A crucial ingredient for the efficiency of the scheme is a random model of the electric conductivity based on the Karhunen-Lo\`{e}ve expansion. It requires the computation of the eigenvalues and eigenfunctions of the covariance matrix at a discrete set of frequency points. An exponential decay of the eigenvalues was observed, allowing for a significant reduction of the number of random parameters. Additionally, the new parameters are uncorrelated. Considering uncertainties in the conductivity and permittivity at the same time requires the study of cross-correlation effects which is the subject of ongoing work.

\section*{Acknowledgment}
U. R\"omer and S. Sch\"ops acknowledge the support of the DFG through the Graduate School of Computational Engineering in Darmstadt.


\begin{thebibliography}{1}

\bibitem{gabriel2009}
C.~Gabriel, A.~Peyman, and E.~H. Grant, ``Electrical conductivity of tissue at frequencies below 1 MHz,'' \emph{Phys. Med. Biol.}, vol. 54, pp. 4863--4878, 2009.

\bibitem{schmidtieee2013}
C.~Schmidt, P.~Grant, M.~Lowery, U.~van Rienen, ``Influence of Uncertainties in the Material Properties of Brain Tissue on the Probabilistic Volume of Tissue Activated,'' \emph{IEEE Trans. Biomed. Eng.}, vol. 60, pp. 1378-1387, 2013. 

\bibitem{mcintyre2002}
C.~McIntyre, A.~Richardson, W.~Grill, ``Modeling the excitability of mammalian nerve fibers: influence of afterpotentials on the recovery cycle.,'' \emph{J. Neurophysiol}, vol. 87, pp. 995--1006, 2002. 

\bibitem{loeve1978}
M.~Lo\`{e}ve, \emph{Probability Theory}, Springer, New York, 1978.

\bibitem{ghanem1991}
R.~Ghanem, P.~D. Spanos, \emph{Stochastic Finite Elements: a Spectral Approach}, Springer, New York, 1991.

\bibitem{xiu2005}
D.~Xiu, and Jan~S. Hesthaven, ``High-order collocation methods for differential equations with random inputs,'' \emph{SIAM Journal on Scientific Computing}, vol. 27.3, pp. 1118--1139, 2005.

\bibitem{babuska2010}
I.~Babu\v{s}ka, F.~Nobile, and R.~Tempone, ``A stochastic collocation method for elliptic partial differential equations with random input data,'' \emph{SIAM review}, vol. 52.2, pp. 317--355, 2010.

\bibitem{elia2013coarse}
M.~D'Elia, and M.~Gunzburger, ``Coarse-Grid Sampling Interpolatory Methods for Approximating Gaussian Random Fields,'' \emph{SIAM/ASA J. Uncertainty Quantification}, vol. 1, pp. 270--296, 2013.  


\bibitem{roemer2016}
U.~R\"omer, S.~Sch\"ops, T.~Weiland, ``Stochastic Modeling and Regularity of the Nonlinear Elliptic curl--curl Equation,'' \emph{SIAM/ASA Journal on Uncertainty Quantification}, vol. 4, pp. 952-979, 2016. 

\bibitem{butson2006}
C.~R. Butson, and C.~C. McIntyre, ``Tissue and electrode capacitance reduce neural activation volumes during deep brain stimulation,'' \emph{Clinical neurophysiology}, vol. 116, pp. 2490--2500, 2005. 

\bibitem{novak1999}
E.~Novak, and K.~Ritter, ``Simple cubature formulas with high polynomial exactness,'' \emph{Constructive approximation}, vol. 15.4, pp. 499--522, 1999.

\bibitem{grant2010}
P.~F. Grant, and M.~M. Lowery, ``Effect of Dispersive Conductivity and Permittivity in Volume Conductor Models of Deep Brain Stimulation,'' \emph{IEEE Transactions on Biomedical Engineering}, vol. 57, pp. 2386--2393, 2010. 

\end{thebibliography}
\end{document}